\documentclass[twocolumn]{aastex6}
\usepackage{natbib}
\begin{document}

\shorttitle{Halo migrants in APOGEE}
\shortauthors{Martell et al.}
\title{Chemical tagging in the SDSS-III/APOGEE survey: new identifications of halo stars with globular cluster origins}
\author{Sarah L. Martell\altaffilmark{1}}
\affil{School of Physics\\
University of New South Wales\\
Sydney, Australia 2052}
\email{s.martell@unsw.edu.au}

\author{Matthew D. Shetrone}
\affil{McDonald Observatory\\
University of Texas at Austin\\
McDonald Observatory, TX 79734, USA},

\author{Sara Lucatello}
\affil{INAF-Osservatorio Astronomico di Padova\\
Vicolo dell'Osservatorio 5\\
35122, Padova, Italy}

\author{Ricardo P. Schiavon}
\affil{Liverpool John Moores University\\
Liverpool, L3 5RF, UK}

\author{Szabolcs M\'esz\'aros}
\affil{ELTE Gothard Astrophysical Observatory\\
Szent Imre Herceg St 112\\
H-9704 Szombathely, Hungary}

\author{Carlos Allende Prieto}
\affil{Instituto de Astrofísica de Canarias (IAC)\\
Vía Lactea s/n\\
E-38200 La Laguna, Tenerife, Spain}

\author{D. A. Garc\'ia Hern\'andez}
\affil{Instituto de Astrofísica de Canarias (IAC)\\
Vía Lactea s/n\\
E-38200 La Laguna, Tenerife, Spain\\
and\\
Departamento de Astrofísica\\
Universidad de La Laguna (ULL)\\
E-38206 La Laguna, Tenerife, Spain}

\author{Timothy C. Beers}
\affil{Department of Physics\\
and\\
JINA Center for the Evolution of the Elements\\
University of Notre Dame\\
Notre Dame, IN 46556, USA}

\and

\author{David L Nidever}
\affil{Large Synoptic Survey Telescope\\
950 North Cherry Ave\\
Tuscon, AZ 85719, USA\\
and\\
Steward Observatory\\
933 North Cherry Ave\\
Tuscon, AZ 85719, USA}

\altaffiltext{1}{s.martell@unsw.edu.au}

\begin{abstract}
We present new identifications of five red giant stars in the Galactic halo with chemical abundance patterns that indicate they originally formed in globular clusters. Using data from the Apache Point Observatory Galactic Evolution Experiment (APOGEE) Survey available through Sloan Digital Sky Survey Data Release 12 (DR12), we first identify likely halo giants, and then search those for the well-known chemical tags associated with globular clusters, specifically enrichment in nitrogen and aluminum. We find that $2\%$ of the halo giants in our sample have this chemical signature, in agreement with previous results. Following the interpretation in our previous work on this topic, this would imply that at least $13\%$ of halo stars originally formed in globular clusters. Recent developments in the theoretical understanding of globular cluster formation raise questions about that interpretation, and we concede the possibility that these migrants represent a small fraction of the halo field. There are roughly as many stars with the chemical tags of globular clusters in the halo field as there are in globular clusters, whether or not they are accompanied by a much larger chemically untaggable population of former globular cluster stars. 
\end{abstract}


\section{Introduction}
The formation process for the stellar halos of disk galaxies is a complex and unsolved problem. Stars from the earliest events in hierarchical assembly should be found throughout the Galaxy (e.g., \citealt{BK07};  \citealt{T10}), while stars from later accretion of lower-mass galaxies mainly orbit in the outer halo, where streams of debris are long-lived thanks to long dynamical timescales. Indeed, the spatial coherence of merger debris can be observed in deep imaging of many galaxies (eg, \citealt{MC80}; \citealt{MD15}), and kinematic coherence has been identified through spectroscopic studies in the Milky Way (e.g., \citealt{SR09}) and M31 (e.g., \citealt{GG09}). Minor mergers clearly play a crucial role in assembling the outer halo, with its high degree of substructure and small or negative net rotation (e.g., \citealt{PV14}; \citealt{DB12}; \citealt{FM11}). However, the inner halo, which is distinguishable from the outer halo both kinematically and chemically (e.g., \citealt{CB07}; \citealt{CBC10}; \citealt{HY12}), may have formed a non-negligible fraction of its mass {\it in situ} (e.g., \citealt{TB14}; \citealt{CM13}) from gas accreted by the Milky Way at early times.

The site and process for the formation of globular clusters are also unclear. Integrated-light studies of extragalactic globular cluster systems (e.g., \citealt{BS06} and references therein) tend to divide them into two broad families: these are variously described as blue and red, old and young, or native and accreted. This is also consistent with the globular clusters in the nearest Local Group galaxies, in which we can study individual stars. In the Milky Way, the age-metallicity relation of globular clusters has two branches with different spatial distributions \citep{MF09} that correspond to these two families. One of the most dramatic results from the PAndAS survey of M31 was the discovery by \cite{MH10} that many of the globular clusters in the outskirts of the galaxy are spatially coincident with tidal streams. Work is ongoing to confirm this association kinematically (e.g., \citealt{ML14}), but the imaging data clearly suggest that the globular clusters in the outer halo of M31 were captured along with the dwarf galaxies in which they originally formed. 

We can investigate the origins of individual stars using chemical tagging, which is based on the principle that the chemical abundance patterns of stars reflect the site of their formation (eg, \citealt{FBH02}). There are many ways to use chemical tagging at different levels of detail, from a coarse disk vs halo separation based on metallicity (e.g., \citealt{SS51}) to membership selection for moving groups (e.g., \citealt{DS07}) to a high-precision search for stars that formed with the Sun (e.g., \citealt{RB14}). \cite{HC16} have recently demonstrated that chemical tagging in many elemental abundances simultaneously can be used to identify known star clusters and the Sagittarius dwarf galaxy in a large, homogeneous data set.

The identification of field stars that originally formed in globular clusters is an application of chemical tagging with direct bearing on the origin of the {\it in situ} component of the Galactic halo. It is possible because globular clusters appear to be the only astrophysical environment to imprint light-element anticorrelations on a fraction of their stars at all evolutionary phases (e.g., \citealt{K79}; \citealt{HB80}; \citealt{CB09}; \citealt{GH15}). The basic pattern is depletion in C, O and Mg simultaneous with enrichment in N, Na and Al, but the extent of the enhancements and depletions varies from cluster to cluster (e.g., \citealt{CB10}; \citealt{MM15}), and these anticorrelations are sometimes joined by variations in the abundances of F \citep{DO13}, Si \citep{YR14} and r-process elements \citep{MS11}. For consistency with previous literature, we refer to this pattern as the "characteristic globular cluster abundance pattern", with stars that have abundance patterns similar to field stars called "first-generation" and stars exhibiting anticorrelated enhancements and depletions called "second-generation".

\subsection{Globular cluster formation and dissolution models}
Although this characteristic pattern can be found in nearly every globular cluster in the Milky Way (e.g., \citealt{CB09}; \citealt{VG13}), there has not yet been a model put forward that completely explains the origin of these abundance anomalies, or why they appear to originate only in globular clusters. The ratio of first- to second-generation stars, the correlations between the extent of abundance variations and present-day cluster properties, and the phase-space distributions of those populations, ought to provide strong constraints on models for globular cluster formation and self-enrichment. Because the pattern resembles the result of high-temperature hydrogen fusion cycles, schematic models have been developed in which globular clusters contain two separate generations of stars, with the second chemically influenced by feedback from the first (e.g., \citealt{CS11}; \citealt{DE10}; \citealt{DM09}; \citealt{DM07}).  These two-generation models immediately encounter a serious problem with the "mass budget" - that is, there are roughly as many chemically unusual stars as chemically normal stars in globular clusters. Since chemical feedback from a star is unlikely to be more than a few percent of its mass \citep{CD91} and star formation is generally inefficient \citep{LL03}, the original generation of chemically normal stars must have been quite large in order to generate enough mass in chemically-enriched winds to produce the abundance variations observed in second-generation globular cluster stars. However, in order for this to be true, the majority of these first-generation stars must have left the cluster following the second episode of star formation, leaving behind the roughly even ratio between the populations that is observed today. It is difficult to imagine a mechanism for rapid loss of at least $90\%$ of the cluster's mass that would not cause total dissolution. Type II supernovae have been suggested as a way to remove any remaining gas, flatten the gravitational potential and free first-generation field stars at large cluster radii, but to date there has been no numerical modelling to verify that this process would work as envisioned. The mass budget problem becomes a crisis in environments like the Fornax dwarf galaxy \citep{LS12} and the inner Milky Way \citep{S16}, where two-generation models predict that the number of stars that must have escaped from globular clusters is larger than the number of stars at globular cluster-like metallicity in the field. 

Self-enriching globular cluster formation models also do not sufficiently explain how the material to construct the second generation manages to stay gravitationally bound to the protocluster, how two short bursts of star formation and a phase of self-enrichment can happen before any supernova enrichment, or how it has happened that no single-generation globular clusters were formed that have survived to the present day. This last point is a natural feature of the model proposed by \citet{K15}, who calculate a minimum mass for star cluster survival to the present day that depends on internally and externally driven mass loss processes, which (for clusters in the Milky Way) is quite similar to the minimum mass for cluster self-enrichment. This problem is difficult to approach with n-body simulations, since it involves high densities, short timescales, magnetohydrodynamics, gas physics and kinetic and radiative feedback. In addition, the observational cataloguing of the phenomenon is fairly complete (e.g., \citealt{CB09}; \citealt{MM15}), with more precise analysis uncovering further complexity (e.g., \citealt{PM15}). Alternative models have been proposed that allow supernovae to participate in cluster chemical evolution without significant effects on the metallicity distribution \citep{S10} or produce multiple abundance populations through mergers of protoclusters \citep{CB10b}. A model by \cite{BL13} suggests that accretion of material from the winds of AGB stars and massive binaries in globular clusters onto the protostellar disks of young stars in those clusters can imprint light-element abundance variations without a second generation of star formation. Stars on different orbits will spend more or less time passing through the cluster center, where the accretable material will be most concentrated, producing the observed range in the strength of abundance variations. Recently there has been new effort made to model the process of star cluster formation (e.g., \citealt{BC15}), and the early results from those studies underline that none of the existing models can reproduce all of the observational aspects simultaneously.

\subsection{Chemical tagging of migrant field stars}
Regardless of the origin of the characteristic globular cluster abundance pattern, its apparent uniqueness makes it a useful chemical tag for identifying stars that formed in globular clusters and have since migrated into other components of the Galaxy. A small number of studies have done exactly this, searching large collections of spectroscopic data for field stars that follow the globular cluster abundance pattern. \cite{MG10} and \cite{M11} used the SDSS-II/SEGUE \citep{Y09} and SDSS-III/SEGUE-2 \citep{E11} surveys, respectively, as their data sources, while \citet{RM12} and \citet{CB10} used literature compilations originally assembled for other purposes, and \citet{LK15} used the Gaia-ESO Survey \citep{GR12}. Due to differences in the data available to the various authors, a number of different chemical tags have been used to identify these migrant stars, but all involve some part of the characteristic globular cluster abundance pattern.

In all of those studies the basic interpretation has been consistent: based on their abundance patterns, these stars must have formed within globular clusters. The fraction of halo field stars chemically tagged as migrants from globular clusters is a few percent, and some authors (e.g., \citealt{MG10}, \citealt{M11}) assume that these stars signal the complete disruption of globular clusters, while others (e.g., \citealt{SC11}; \citealt{LK15}) assume that only a small proportion of stars escape from the globular clusters in which they formed. A more thorough consideration of the initial properties and orbits of globular clusters, and their stability against various mass-loss processes in an evolving galactic potential, would be needed to know which of these interpretations is correct. 

\begin{figure*}[thb!]
\plottwo{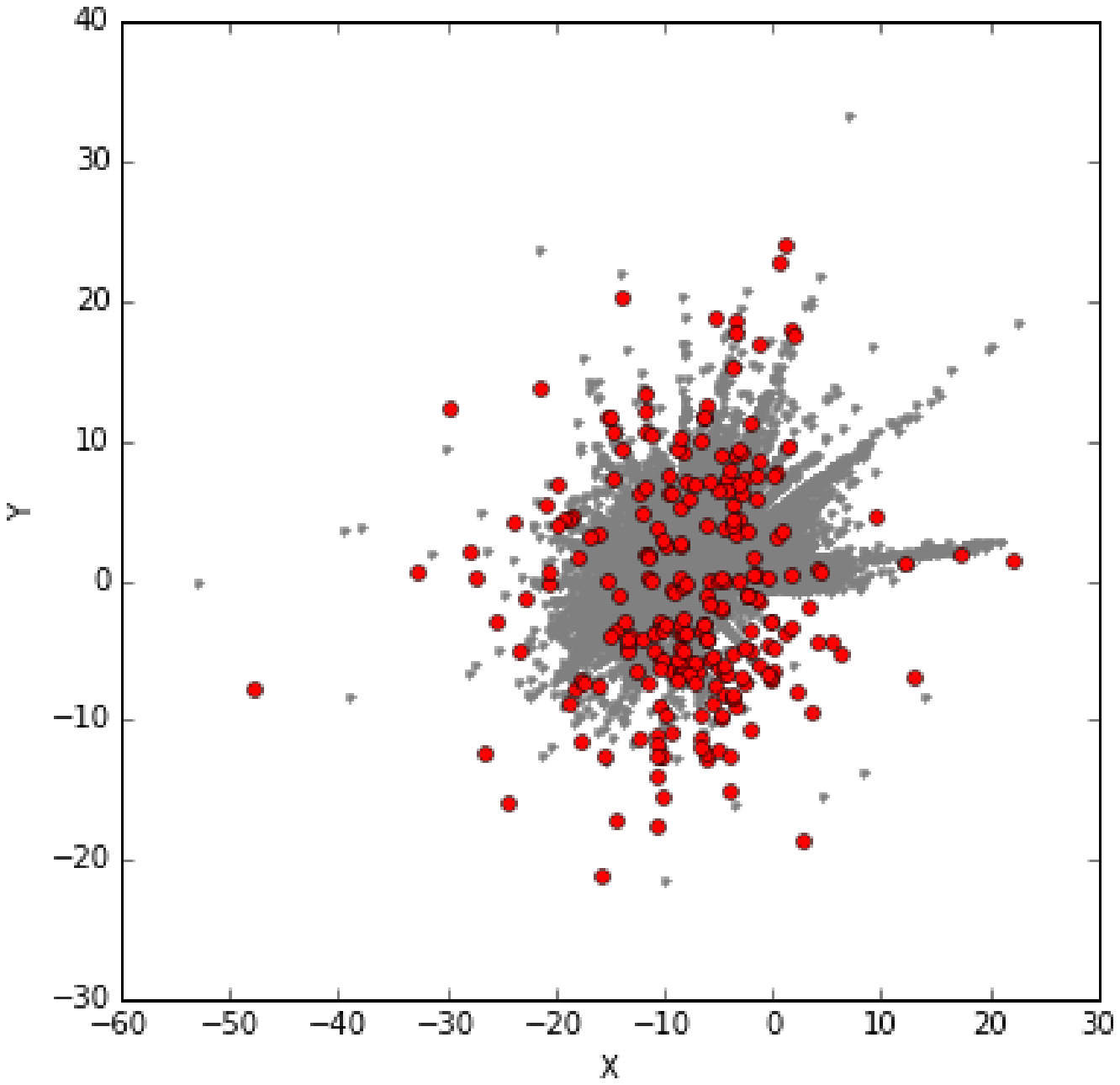}{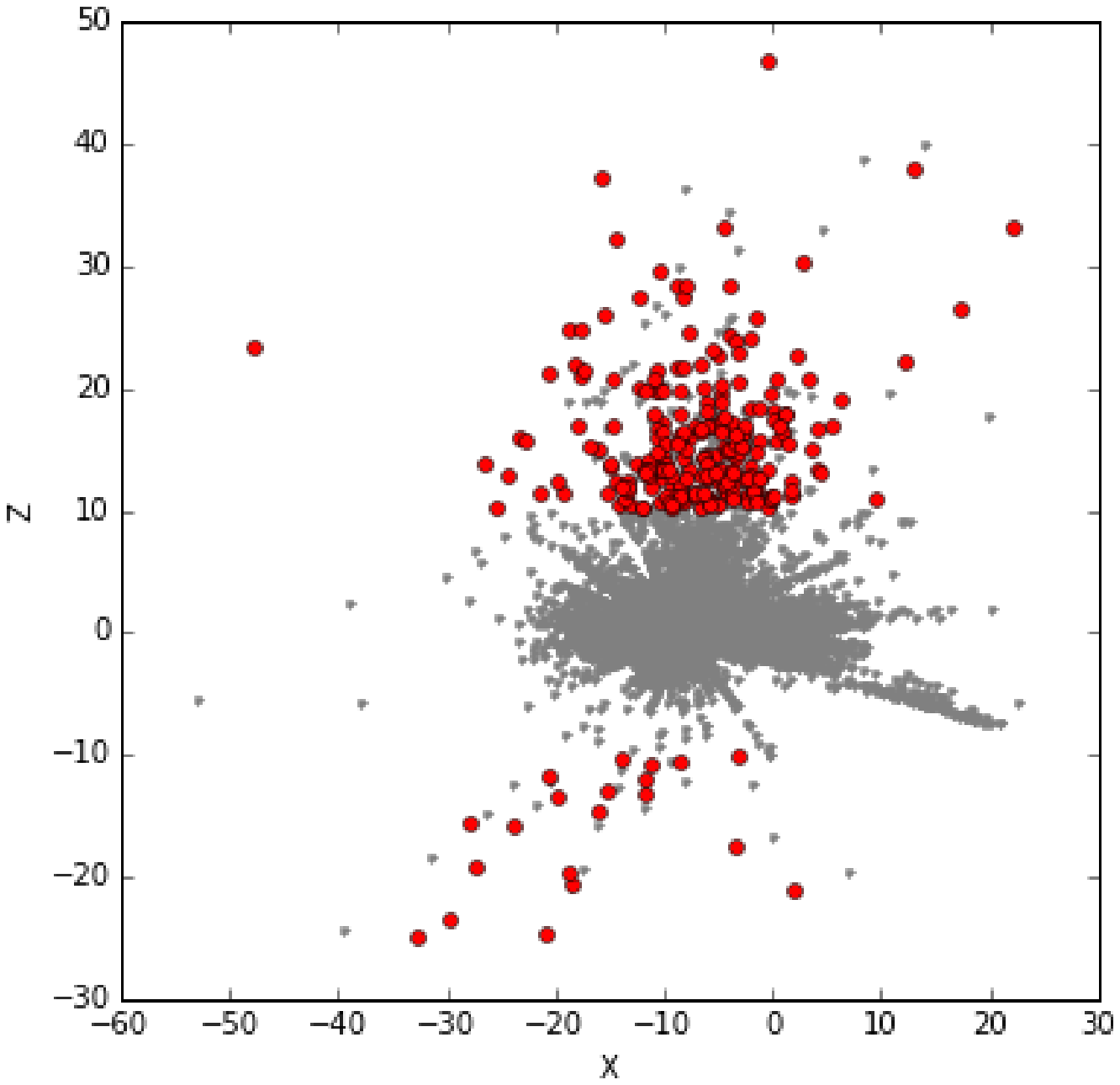}
\caption{Galactic XYZ coordinate locations for the high-quality data set (small grey points) and the final data set (red circles). The requirement that stars be at least 10 kpc from the plane of the Galaxy can be seen; beyond that limit, the distribution of halo RGB stars is similar to that of the general population.}
\end{figure*}

New identifications of halo field stars that can be chemically tagged to globular clusters as their formation site allow us to reconsider the question of {\it in situ} halo formation with an expanded data set. In this publication we discuss 253 halo red giant branch (RGB) stars from the Apache Point Observatory Galactic Evolution Experiment (APOGEE) survey data set, including five that can be chemically tagged back to globular clusters. These results emphasise the central importance of large-scale Galactic archaeology surveys as a way to identify rare objects in the Galaxy, and as a way to investigate the general process of galaxy formation using the Milky Way as a proxy for spiral galaxies in general.

\section{The data set}
As with any search for unusual objects, this study requires a large data set. The results reported in this paper are based on stellar parameters and elemental abundances for the 156593 unique stars from Data Release 12 \citep{DR12} of the APOGEE survey \citep{M15}. One of four Sloan Digital Sky Survey-III (SDSS-III, \citealt{E11}) experiments, APOGEE used a high-resolution spectrograph on the Sloan 2.5 m telescope \citep{GS06} to obtain H-band spectra (R=22,500) for stars distributed across all Galactic components. Precision radial velocities, stellar parameters, and abundances for up to 15 elements have been obtained from these spectra. Further detail on the APOGEE survey goals, observations, data, and the data reduction pipeline can be found in \citet{M15}, \citet{ZJ13}, \citet{HS15}, and \citet{NH15}, respectively. The APOGEE Stellar Parameters and Chemical Abundances Pipeline (ASPCAP) is described in detail in \citet{GP15} and chooses a best fit result based on a pre-computed grid of stellar spectra \citep{Z15}.

We first select for valid data using flags set during the data reduction and analysis process\footnote{These are described online at http://www.sdss.org/dr12/algorithms/bitmasks/}: at this stage 20605 stars with aspcapflag 'STAR\_BAD', which is a catch-all indicator of trouble, noting that either the data quality is low, or that at least one of the derived quantities T$_{\rm eff}$, log(g) is outside bounds, or that the star has a large apparent rotational velocity, were rejected. We then eliminate known globular cluster stars from the sample: 3060 stars with targflags 'APOGEE\_SCI\_CLUSTER' or 'APOGEE\_CALIB\_CLUSTER' were rejected, as were a further 31 stars which are likely to be serendipitously observed globular cluster members in the APOGEE data set (Shetrone, priv. comm.). 

The next step is to require a certain level of quality in data and in analysis results: mean signal-to-noise ratio per half-resolution element (the APOGEE spectroscopic figure of merit) was required to be at least 80, and log(g), T$_{\rm eff}$, metallicity, and nitrogen and aluminium abundances and distance d (calculated based on isochrones as described in \citealt{HH14}) were required to have valid values (i.e., not 9999 bad-value placeholders). This step is taken using the uncalibrated DR12 stellar parameters and the calibrated abundances, as described in \citet{HS15}, and reduces the sample to 87252 stars. 

Finally, to select likely members of the halo we use a combination of surface gravity, metallicity, effective temperature and height above the Galactic plane: log(g) $< 3.0$, $-1.8 <$ [Fe/H]$< -1.0$, T$_{\rm eff} < 4500$K, $\mid{\rm z}\mid> 10$ kpc. This returns a final data set of 253 likely halo giants with high-quality spectra and reliable parameters and abundances. As before, we use the uncalibrated stellar parameters for log(g) and T$_{\rm eff}$, and the calibrated abundance for [Fe/H]. The vertical coordinate $\mid{\rm z}\mid$ is the absolute value of d sin(b), where b is the Galactic latitude.

This selection may introduce biases, though it is difficult to state conclusively what the overall effect on the data set is since the biases operate in independent directions. By requiring SNR of at least 80 we are preferentially selecting giants with lower surface gravity as distance from the Sun increases. Most of the light elements that can be used for chemical tagging of globular cluster migrants are not affected by surface gravity. However, nitrogen is affected by the typical evolution in surface abundances experienced by stars as they ascend the red giant branch (e.g., \citealt{MSB08}; \citealt{GSCB00}), in the sense that more evolved stars (at a fixed metallicity and mass) tend to have higher nitrogen abundances. Nitrogen abundance does not show any clear trend with distance from the Sun for stars in our final data set with a narrow range in SNR (from 120 to 150) and metallicity (from $-1.7$ to $-1.4$), indicating that this particular selection bias does not have a strong effect on our results. 

The lower limit on metallicity is imposed because the nitrogen abundances are unreliable for lower-metallicity stars \citep{MM15}. By requiring metallicity below $-1.0$ we reject the majority of stars in the thin disk, and by requiring a height of at least 10 kpc above the plane, we avoid the majority of the thick disk but also reject halo stars currently within 10 kpc of the plane. In this metallicity range, it is difficult to identify halo stars in the Solar neighborhood with confidence without kinematic information. Requiring a metallicity of at least $-1.8$ makes our target stars more likely to belong to the inner-halo population than the outer-halo population, though there is not a clear dividing line between the two (e.g., \citealt{CB07}; \citealt{CBC10}). Fortunately, the magnitude of the light-element abundance pattern in globular clusters is not a dramatic function of metallicity in the range we select (e.g., \citealt{MM15}). This metallicity range is the same as was used in \citet{MG10} and \citet{M11}, but this is coincidental: in the earlier work it ensured that the CN molecular absorption feature with a bandhead at $3883\hbox{\AA}$ was sensitive enough to changes in nitrogen abundance, and in the current work it ensures that the APOGEE nitrogen abundances, based on different spectral features, are reliable.

We compare our final data set to the 87252 stars with high-quality APOGEE spectra and parameters through a series of figures: Figure 1 shows Galactic (X,Y) and (X,Z) distributions, with the high-quality data set shown as smaller grey points and the final data set drawn as filled red circles. Figure 2 shows the metallicity distribution function (MDF) for the high-quality data set (solid line) and the final data set (dotted line). The final data set appears to follow the distribution of the high-quality data set more than 10 kpc from the Galactic plane, and the MDF for the final data set does not have a significantly different shape from the MDF for the high-quality data set in the metallicity range in which they overlap.

\begin{figure}[ht!]
\plotone{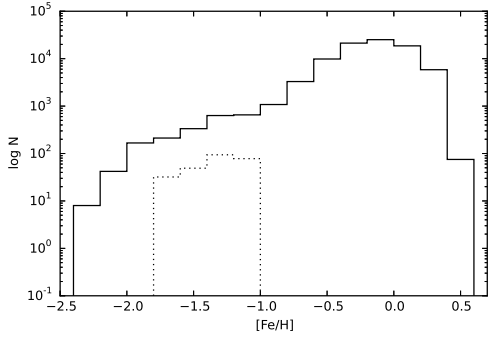}
\caption{Logarithmic metallicity distribution functions for the high-quality data set (solid line) and the final data set (dotted line). The final data set does not differ strongly from the high-quality data set within its restricted metallicity range, indicating that our selection is not introducing any problematic biases in metallicity.}
\end{figure}

\section{Chemical tagging in APOGEE data}
Ideally, chemical tagging uses as many elements as will contribute new information; that is, abundances carrying redundant information can be omitted without a loss of confidence in the result. There is some apparent decoupling between the carbon-nitrogen anticorrelation and the oxygen-sodium anticorrelation in globular clusters (e.g., \citealt{S15}), presumably driven by differences in data characteristics and analysis techniques and by the fact that the nuclear reaction chains that cause these anticorrelations occur at different temperatures and possibly in different stars entirely. As a result, we evaluated all of the light elements that typically participate in the characteristic globular cluster abundance pattern (carbon through aluminum) for usefulness in this study. Unfortunately, sodium and oxygen, which would make an excellent comparison to the literature because they are so often used for studying multiple populations in globular clusters using high-resolution optical spectra, are not useful for this data set because the sodium lines in APOGEE spectra are too weak in this metallicity range to be reliably used for abundance analysis, and the ASPCAP oxygen abundances for oxygen-poor stars are known to be too large owing to a degeneracy between temperature and [O/Fe]. 

Of the four remaining abundances (C, N, Mg and Al), nitrogen and aluminum appear to be the most effective as chemical tags for globular cluster-like abundance patterns. The panels of Figure 3 show the distribution of [C/Fe], [N/Fe], [Mg/Fe] and [Al/Fe] versus effective temperature for our final data set, with APOGEE DR12 data for stars in M3 and M13 (which occupy the same metallicity range as our final data set) overplotted as filled blue circles and purple squares, respectively. There is clear variation in the nitrogen and aluminum abundances in the cluster data, and also in the field stars. The carbon and magnesium abundances in cluster stars do not show much range, similar to the measurements reported by \citet{CB09}, although there are notably carbon-rich stars in the field.

\begin{figure}[ht!]
\plotone{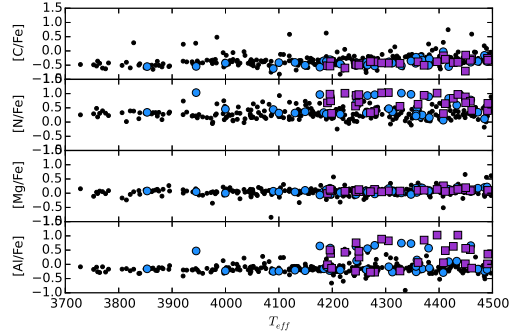}
\caption{Abundances of carbon, nitrogen, magnesium and aluminum for stars in the final data set (small black points) and for stars in the globular clusters M3 (blue circles) and M13 (purple squares). While [N/Fe] and [Al/Fe] show a clear range of abundance in both sets of stars, the range in [C/Fe] and [Mg/Fe] is more compressed in the globular cluster stars.}
\end{figure}

Our selection for globular cluster migrants in the halo field begins with a nitrogen- and metallicity-based selection criterion similar to the one used in \citet{S16}. Using over 5000 RGB stars within 3 kpc of the Galactic center, they fit a sixth-order polynomial to the distribution in the [N/Fe]-[Fe/H] plane and select all stars more than 4$\sigma$ above that curve as nitrogen-rich. With a more limited metallicity range, we find that a third-order polynomical captures the mean behavior of our data set well. We label all stars with nitrogen abundance more than 0.335 dex above the mean at fixed metallicity as "nitrogen-rich", which is the same as the selection in \cite{S16}. This returns seven stars that are nitrogen-rich relative to the final data set. Figure 4 shows our final data set in the [N/Fe]-[Fe/H] plane, with the selection criterion shown as a dashed line and the seven nitrogen-rich stars plotted as red triangles.

\begin{figure}[ht!]
\plotone{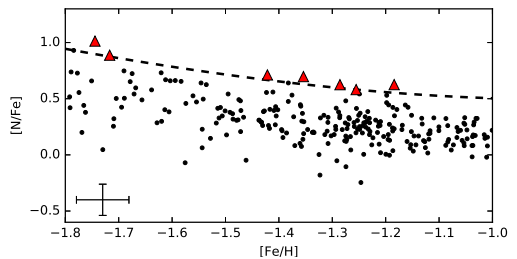}
\caption{Nitrogen abundance versus overall metallicity for the 253 stars in our final data set, shown as small black points. The dashed line represents the selection criterion for nitrogen-rich stars described in the text, and the seven stars that lie above that curve are drawn as red triangles.}
\end{figure}

However, an enhanced nitrogen abundance is not by itself a sufficient indicator of a globular cluster-like abundance pattern. In \citet{MG10} and \citet{M11}, depletions in carbon abundance were also required for successful chemical tagging, and \citet{S16} remove stars that are rich in both carbon and nitrogen from their sample, since their surface abundances may have been modified by mass transfer from a low-mass companion asymptotic giant branch (AGB) star. In order to confidently state chemical tagging results for this data set, we require enhanced nitrogen, non-enhanced carbon (similar to the \citealt{S16} selection) and enhanced aluminum abundance. 

One of the seven nitrogen-rich stars in our final data set has a carbon abundance higher than a reasonable CH-star limit of $+0.3$, leaving six candidates. The carbon-rich star, 2MASS J15015733+2713595, shows a distinct difference between the heliocentric radial velocities measured from its first two observations and the next three, which took place 300 days later. This radial velocity data, given in Table 1, is consistent with this star having a binary companion. Long term radial velocity monitoring of all N- and Al-rich stars would naturally be the best course to establish the number of such objects formed through the binary channel. However, very few of the N-rich stars in our sample have been observed with a baseline of more than a few months and none of them have been regularly monitored, limiting the orbital properties than can be detected in this data set. None of the six N-rich candidates without strong carbon enhancement has a particularly strong variability in its radial velocity over the period of the APOGEE observations.

\begin{table}[b!]
\centering
\caption{Radial velocities for individual APOGEE observations of the nitrogen-rich, carbon-rich star identified in this study}
\begin{tabular}{lcr}
\tablewidth{0pt}
\hline
\hline
APOGEE ID & Observation MJD & v$_{\rm helio}$ (km s$^{-1}$)\\
\hline
2M15015733+2713595 & 56408 & -121.359\\
2M15015733+2713595 & 56431 & -125.403\\
2M15015733+2713595 & 56706 & -104.621\\
2M15015733+2713595 & 56724 & -106.327\\
2M15015733+2713595 & 56733 & -107.549\\
\hline
\end{tabular}
\end{table}

Binary mass transfer from an intermediate-mass AGB star can cause high surface abundances of nitrogen and aluminum in its companion. Intermediate-mass AGB stars (M$\sim$3-8\,M$_{\odot}$) undergo Hot Bottom Burning, producing significant amounts of N and in some cases Al \citep{ventura13}. This material can be transferred onto a binary companion if the orbital separation is within the correct range. The companion star's atmosphere acquires the signature of AGB nucleosynthesis, which persists to the present day, when the donor star continues evolving, eventually becoming a faint white dwarf.

As discussed in \citet{S16}, establishing the expected number of those stars in a given population from first principles is rather uncertain, requiring knowledge of a number of properties of the population, such as the initial mass function, the binary fraction, and the distributions of period, eccentricity and mass ratio. An alternative approach to deriving the expected number of N- and Al-rich stars of binary origin is that of scaling, according to the IMF, the number of CH-stars in the sample. Just like for Ba and CEMP-s stars, the peculiar composition of CH-stars is due to mass transfer from a low mass (M$\sim$1.5-4\,M$_{\odot}$)\footnote{The mass range for the donor star is determined by the minimum mass for the third dredge-up (and hence the minimum mass for becoming C-rich) and by the onset of Hot Bottom Burning.} AGB companion \citep[see e.g.][]{mcclure90,lucatello05,starkenburg14}. 

If we assume that the existence of a binary companion, and the distributions of orbital periods, mass ratios and eccentricity, are not dependent on the primary star's mass (which is quite reasonable in the mass range of the dataset under consideration), the difference between the expected incidence of CH-stars and N- and Al-rich stars should only depend on the incidence of donor stars in a given population, that is to say the frequency of objects in the 1.5-3\,M$_{\odot}$ (companions to the CH-stars) and 3-8\,M$_{\odot}$ (companions to the N and Al rich stars) mass ranges. The ratio between the number of stars in these two mass ranges is rather similar, regardless of which of the most commonly adopted IMFs \citet{salpeter55,kroupa01} or \citet{chabrier03} is used, and it is of order $\sim 0.5$. In a given population, then, we expect half as many N- and Al-rich stars as CH-stars. We note that this is an aggressive estimate: all stars in the $\sim$3-8\,M$_{\odot}$ range will undergo hot bottom burning and become N-rich, but the MgAl cycle only operates at temperatures above T$\sim$50\,MK \citep[][]{ventura13}, so that not all intermediate-mass AGB stars will produce significant amounts of aluminum. 

On the basis of this ratio and of the number of {\it bona fide} CH stars, we can estimate the expected number of {\it bona fide} N-rich stars of binary origin. Six of the stars in the final data set have a reliably measured C abundance above a reasonable CH star threshold of [C/Fe]$>+$0.3~dex, indicating that as many as as three of the N- and Al-rich stars in the final data set may owe their atmospheric composition to binary mass transfer. We emphasise that this is likely an overestimate, based on adopting a wide mass range for the production of Al in intermediate-mass stars (see \citealt{S16} for further discussion on this point).

To determine the level of aluminum enrichment that would effectively tag stars as migrants from globular clusters, we compare our six candidates (omitting the carbon-rich star) to red giant stars in M3 and M13, well-studied globular clusters with clear correlations between [N/Fe] and [Al/Fe]. Figure 5 shows this plane, using the same color coding as Fig. 3 for the globular cluster stars. While the majority of stars in our final data set and a fair fraction of the cluster giants lie in a group with slightly supersolar [N/Fe] and slightly subsolar [Al/Fe], the distribution of cluster stars extends to [N/Fe] and [Al/Fe] values around $+1.0$. 

We reject the most nitrogen-rich candidate because it does not show any aluminum enhancement above the mean abundance level of the final data set. However, it is also the most metal-poor of the candidates, and we note the possibility that the APOGEE aluminum abundance may become a lower limit in the metal-poor regime. The remaining five stars fall outside the main group of field stars, and within the envelope of globular cluster stars. We identify these five stars in the positively correlated [N/Fe]-[Al/Fe] tail as new candidates for globular cluster migrants in the halo field, $2\%$ of our sample. Given the uncertainty in the chemical tagging of the most nitrogen-rich star, and the possibility that a few of the N- and Al-rich stars may derive their abundance patterns from AGB mass transfer rather than globular cluster self-enrichment, we conclude that the overall fraction of globular cluster migrants in the final data set is robust, although the chemical tagging of the individual stars may not be.

\begin{figure}[tb!]
\plotone{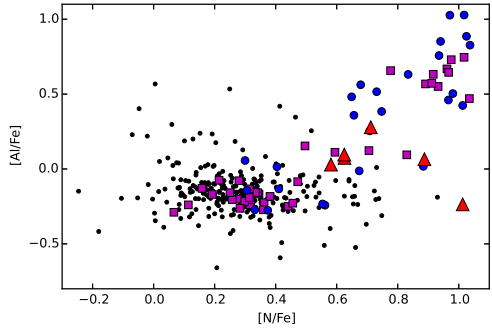}
\caption{Our final data set (small black points) and globular cluster stars (same color coding as in Fig. 3) in the [N/Fe]-[Al/Fe] plane, with the six remaining (non-carbon-rich) candidates drawn as red triangles. The most nitrogen-rich candidate is rejected because {\it is} not more aluminum-rich than the field average, and the five remaining candidates in the extended N-Al abundance distribution are new candidate globular cluster migrants in the field.}
\end{figure}

\begin{table*}[bt!]
\centering
\caption{Nitrogen-rich stars identified in this study}
\begin{tabular}{llcccccccc}
\tablewidth{0pt}
\hline
\hline
APOGEE ID & Classification & T$_{\rm eff}$ & log(g) & [Fe/H] & [C/Fe] & [N/Fe] & [Al/Fe] & R$_{\rm GC}$ (kpc) & $\mid{\rm z}\mid$ (kpc)\\
\hline
2M12555505+4043433 & Migrant & 4070 & 1.01 & -1.42 & -0.39 & 0.71 & 0.28 & 24.21 & 21.19\\
2M15113526+3551140 & Migrant & 4250 & 1.69 & -1.26 & -0.39 & 0.58 & 0.03 & 17.38 & 15.17\\
2M15204588+0055032 & Migrant & 4406 & 1.63 & -1.18 & -0.38 & 0.63 & 0.07 & 13.77 & 13.03\\
2M13251355-0044438 & Migrant & 4585 & 1.47 & -1.72 & -0.05 & 0.89 & 0.06 & 18.65 & 16.31\\
2M17252263+4903137 & Migrant & 4171 & 0.91 & -1.29 & 0.10 & 0.62 & 0.10 & 19.39 & 10.83\\
2M15241679+3536331 & Al-poor & 4442 & 1.57 & -1.74 & -0.45 & 1.01 & -0.24 & 17.23 & 16.32\\
2M15015733+2713595 & C-rich & 4120 & 1.24 & -1.35 & 0.58 & 0.70 & -0.37 & 18.83 & 16.13\\
\hline
\end{tabular}
\end{table*}

\section{Discussion}
This result is a clear confirmation of the results reported in our previous work: a small fraction of the stars in the Galactic halo have light-element abundance patterns that are otherwise only seen in globular cluster stars, and these stars have presumably migrated from clusters into the halo. However, this small fraction does not mean that migrant stars are rare relative to the globular cluster population: to first order, there is as much mass in chemically taggable migrant stars in the field ($2.5\% \times 10^{9}M_{\odot}$, halo fraction from \citealt{MG10} and halo mass from \citealt{FBH02}) as there is in globular cluster stars with second-generation abundances ($67\% \times 160~ {\rm clusters} \times 5\times10^{5}M_{\odot}$ per cluster, second-generation percentage from \citealt{CB10} and globular cluster number and typical mass from \citealt{H96}, 2010 edition).
 
Developing an optimized chemical tag for identifying globular cluster stars in the field would require comparing the effectiveness and false-positive returns of the tags used in studies to date. Ideally, this would be done by measuring all relevant abundances and molecular bandstrengths for all stars claimed to be globular cluster migrants in the halo, comparing against those same quantities for stars currently in globular clusters. While the field versus cluster comparison has been central to our selection of cluster migrants within both SEGUE and APOGEE, the cross-survey comparison is more difficult. Unfortunately, none of the 64 stars discussed in \citet{MG10} and \citet{M11} is among the 80718 stars with high-quality APOGEE spectra and parameters that form the basis of this study.
 
The context for interpretation of this result has changed since our first identification of globular cluster migrants in the halo field. Theoretical studies of halo formation have become more detailed, with finer mass resolution and tracking of some of the fundamental elements of galactic chemical evolution (eg, \citealt{GJ16}; \citealt{TW12}). Practical chemical tagging (eg, \citealt{LJ15}) has also become more sophisticated, using simulated Milky Way-like halos with a history of dwarf galaxy accretion as a testbed for efficient methods of identifying stars from those dwarf systems long after their accretion.

The key development that has upset the scenario presented in our earlier studies has been a renewed effort at modelling globular clusters as multi-generation stellar systems. Fundamentally, there is not a single model that produces complex populations (as described in \citealt{BC15} and \citealt{RD15}), with the variety in abundance behaviour that we see in globular clusters. \citet{K15} presents a model that forms massive clusters at early times and uses minor mergers to scatter them into the halo where they can survive to the present day, but does not investigate the chemical compositions of the stars. We have extensive and excellent observational data on the photometric and abundance complexity in globular clusters (e.g., \citealt{PM15}; \citealt{MS11}; \citealt{CB09}) with which to test new models for globular cluster formation as they are put forward. This topic warrants serious modelling work, to understand how star cluster formation fits into the high-redshift Universe and to explain how the known abundance anomalies arise during that cluster formation process.

As described above, the existing self-enrichment models for globular cluster formation require a large number of first-generation stars that were lost to the field at early times. If this significant first-generation mass loss did occur, then the small fraction of chemically taggable globular cluster stars in the halo represents a much larger migrant population and potentially a significant fraction of the halo. Using the expression from \citet{M11} for ${\rm f}_{\rm h}^{\rm GC}$, the total fraction of halo stars originating in globular clusters, a globular cluster formation scenario with strong early mass loss inflates the $2\%$ of chemically taggable stars we find in this study into $13\%$ of halo stars originating in globular clusters. If an alternative mechanism can produce the characteristic globular cluster abundance pattern without such a dramatic overproduction of first-generation stars, then the ratio of first- to second-generation stars in the halo field should be closer to 1:2, the typical ratio in present-day globular clusters. The total mass contributed to the halo by globular clusters would then be around $4\%$, similar to the total mass still in globular clusters.

Developing a self-consistent, cosmologically situated model for globular cluster formation is crucial for understanding globular clusters as contributors to the Galactic halo. What triggered their short-lived star formation? What fraction of stars formed in this mode remain in clusters today, what fraction of clusters rapidly became unbound to form the general field populations in galaxies, and what fraction of stars in long-lived clusters have escaped in the intervening time? Ultimately, the importance of chemically tagged migrant stars in the Milky Way depends on the site of their formation, and when that site joined the hierarchical merging process that ultimately produced the Galaxy. This invites attention from various angles: a focus on low-mass star clusters in the  Milky Way could give insight on the low-mass limit of the cluster chemical enrichment process; searches for compact star-forming regions at redshift $\sim 3$ might allow us to see the progenitors of today's ancient globular clusters directly; and detailed simulations of star cluster evolution in a realistically evolving galactic potential would clarify which clusters survive their early stages and outline the most important mechanisms for cluster mass loss.

\acknowledgments
Funding for SDSS-III has been provided by the Alfred P. Sloan Foundation, the Participating Institutions, the National Science Foundation, and the U.S. Department of Energy Office of Science. The SDSS-III web site is http://www.sdss3.org/.

SDSS-III is managed by the Astrophysical Research Consortium for the Participating Institutions of the SDSS-III Collaboration including the University of Arizona, the Brazilian Participation Group, Brookhaven National Laboratory, Carnegie Mellon University, University of Florida, the French Participation Group, the German Participation Group, Harvard University, the Instituto de Astrofisica de Canarias, the Michigan State/Notre Dame/JINA Participation Group, Johns Hopkins University, Lawrence Berkeley National Laboratory, Max Planck Institute for Astrophysics, Max Planck Institute for Extraterrestrial Physics, New Mexico State University, New York University, Ohio State University, Pennsylvania State University, University of Portsmouth, Princeton University, the Spanish Participation Group, University of Tokyo, University of Utah, Vanderbilt University, University of Virginia, University of Washington, and Yale University. 

SLM acknowledges financial support from the Australian Research Council through DECRA Fellowship DE140100598. SM has been supported by the J{\'a}nos Bolyai Research Scholarship of the Hungarian Academy of Sciences. TCB acknowledges partial support for this work from grants PHY08-22648; Physics Frontier Center/Joint Institute or Nuclear Astrophysics (JINA), and PHY 14-30152; Physics Frontier Center/JINA Center for the Evolution of the Elements (JINA-CEE), awarded by the US National Science Foundation. D.A.G.H. was funded by the Ram\'on y Cajal fellowship number RYC-2013-14182 and
acknowledges support provided by the Spanish Ministry of Economy and Competitiveness (MINECO) under grant AYA-2014-58082-P. 

\facility{Sloan}
\software{Astropy}

\bibliographystyle{aasjournal}
\bibliography{apm}

\end{document}